\documentstyle[11pt,newpasp,twoside,epsf]{article}
\newcommand{\Teff}{\hbox{$T_{\rm eff}$}}
\def\logg{$\log g$\thinspace}
\def\logm{$\log m$\thinspace}
\newcommand\etal{et al.\,}
\newcommand{\iion}[2]{\mbox{#1\ {\scriptsize #2}}}

\begin{document}
\title{Identification of the 4486,\,4504\AA\ emission lines in O-type spectra}
 \author{K. Werner and T. Rauch}
\affil{Institut f\"ur Astronomie und Astrophysik, Univ.\ T\"ubingen, Germany}

\begin{abstract}
Inspired by an appeal to the community from Walborn (these proceedings) we
decided to solve the long-standing problem concerning the nature of the
4486,\,4504\AA\ emission lines, which are frequently observed in O-type spectra
and which are particularly prominent in supergiants. We claim that these lines
emerge from sulfur, namely by de-excitation of a highly excited \iion{S}{IV}
doublet state. We prove this by an exploratory NLTE calculation with a detailed
model atom.
\end{abstract}

\section{Introduction}

In his paper, Walborn (these proceedings) emphasizes that two prominent
emission lines occurring in O-type spectra remain unidentified 37 years after
being documented by Wolff (1963). In fact, as pointed out by Wolff, these lines
were reported even earlier by Swings \& Struve (1940) in a spectrum they have
taken of 9~Sge, O7.5\,Iaf, hence this problem remains unsolved over 60 years.
Walborn furthermore states that despite several contrary suggestions in the
literature, these lines are unlikely to be due to CNO ions, because they show
no preference for ON or OC spectral types. Because of their possible role as
selective emission lines with potential diagnostic power he concludes that ``we
should not sleep well until these lines are identified''. The prospect of many
sleepless nights, though not too frightening to astronomers, prompted us to
attack this problem.

\begin{figure}[bth]
\epsfxsize=\textwidth
\epsffile{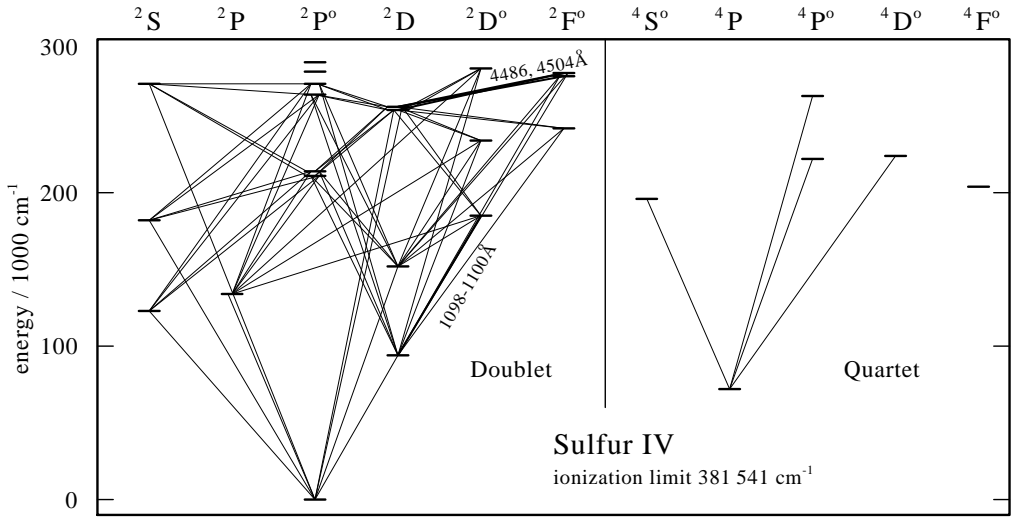}
\caption{Model atom for \iion{S}{IV}. The transitions in the doublet system
responsible for the observed emission are marked by thick lines and wavelength
labels. A FUV line discussed in the text is also marked.}
\end{figure}

\begin{figure}[bth]
\epsfxsize=0.80\textwidth
\epsffile{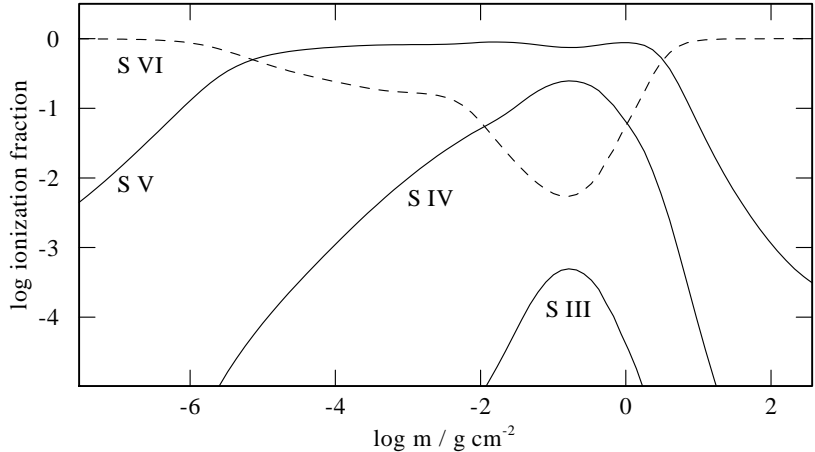}
\caption{Ionization of sulfur as a function of depth in the model atmosphere.
Depth parameter is the column mass measured inward from the top of the
atmosphere.}
\end{figure}

\begin{figure}[bth]
\epsfxsize=\textwidth
\epsffile{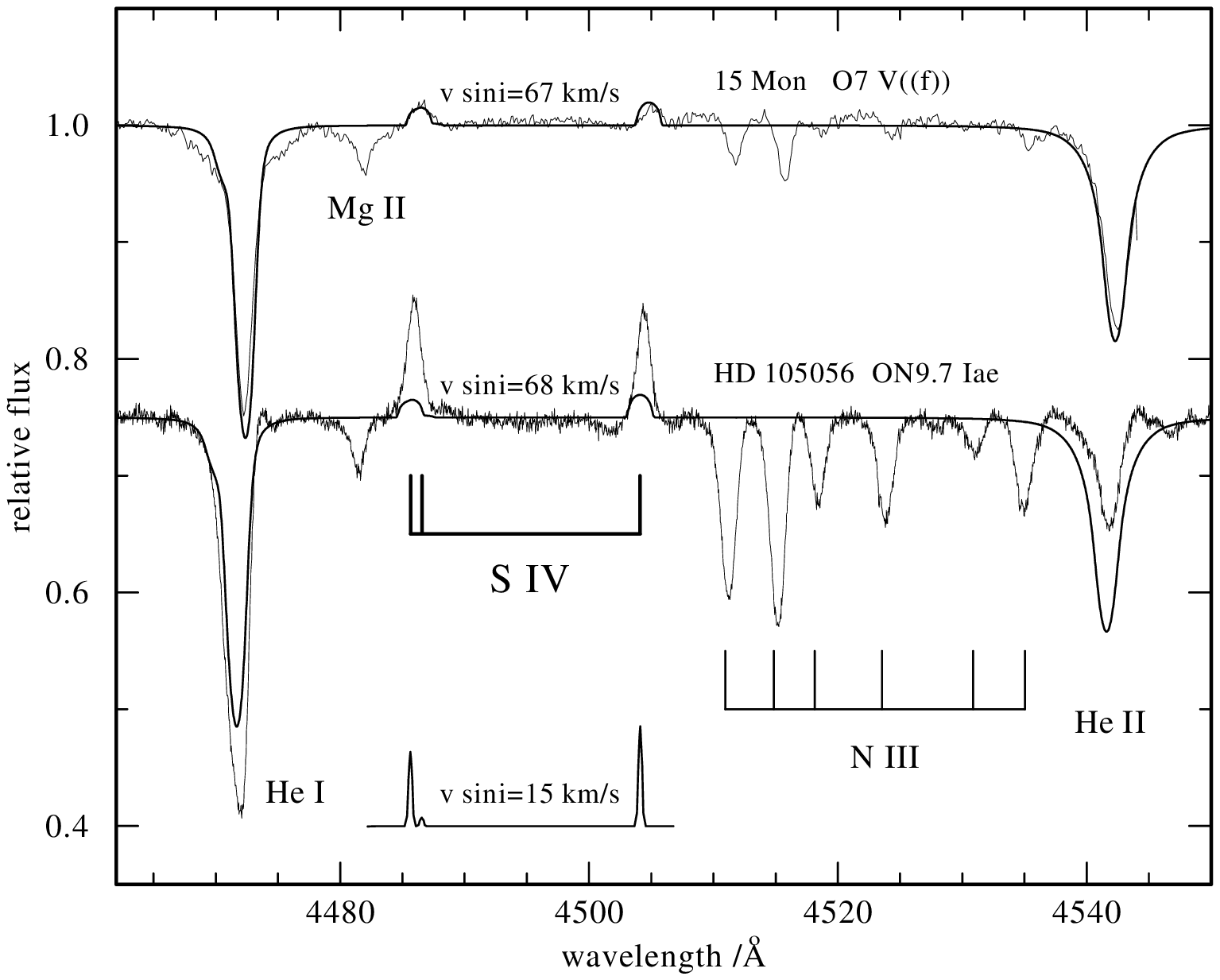}
\caption{Comparison of the synthetic spectrum with 15\,Mon and HD\,105056. The
model is folded according to the stellar rotation velocity and redshifted for
15\,Mon in order to match \iion{He}{I}~4471\AA. The computed spectrum with
reduced rotation velocity and, thus, resolved blue components is shown at the
bottom.}
\end{figure}

\section{NLTE calculations}

We calculated plane-parallel static NLTE model atmospheres composed of
hydrogen, helium, and sulfur. The code used is based on an Accelerated Lambda
Iteration (ALI) technique and is described in detail by Werner \& Dreizler
(1999). The sulfur model atom consists of ionization stages \iion{S}{II--VI}.
Figure~1 displays our model atom for \iion{S}{IV}, which has 27 NLTE levels
linked by 70 radiative transitions. Energy levels were obtained from the NIST
data\-base (http://www.nist.gov/), oscillator strengths and bound-free 
cross-sections from the Opacity Project data\-base (Seaton \etal 1994). Line
broadening and collisional rates are computed by standard approximate
expressions due to the lack of reliable data. In the following we exclusively
describe results from a single model calculation (\Teff=35\,500\,K, \logg=3.7,
solar abundances). Systematic investigations covering the O spectral sequence
and all luminosity classes are the subject of future work.

\noindent
Figure~2 shows the sulfur ionization structure throughout the atmosphere. It
can be seen that \iion{S}{V} is the dominant stage in the line formation region
(\logm= \mbox{$0 ... -0.4$}), followed by \iion{S}{IV}. This situation favors
recombination from  \iion{S}{V} into highly excited \iion{S}{IV} levels which
then de-excite radiatively by photon emission. In fact this turns out to be the
mechanism responsible for the emission lines under discussion. The observed
emission line pair stems from the transition
3s$^2$4d~$^2$D--3s$^2$4f~$^2$F$^{\rm o}$, which in detail is composed of three
lines. In all observed spectra we have inspected in the literature, the two blue
components at 4486\AA\ are too close to be resolved because of stellar rotation
(Fig.\,2).

\begin{table}[bth]
\caption{Atomic data for the newly identified \iion{S}{IV} emission lines,
taken from NIST and Opacity Project databases}
\begin{tabular}{cccccc}
\hline
Wavelength&Rel.&Oscillator& E$_i$ \quad\qquad E$_k$            & J$_i$ -- J$_k$ & g$_i$ -- g$_k$ \\
Air (\AA) &Int.&Strength  & (cm$^{-1})$ \quad (cm$^{-1})$ &                &                \\
\hline
4485.662  & 16 & 0.407    & 255\,395.8 -- 277\,682.8      & 3/2 -- 5/2       & 4 -- 6 \\
4486.568  & 12 & 0.019    & 255\,400.3 -- 277\,682.8      & 5/2 -- 5/2       & 6 -- 6 \\
4504.093  & 14 & 0.388    & 255\,400.3 -- 277\,596.1      & 5/2 -- 7/2       & 6 -- 8 \\
\hline
\end{tabular}
\end{table}

\noindent
Figure~3 compares the computed model spectrum with observations of 15\,Mon,
O7\,V((f)), and HD\,105056, ON9.7\,Iae. The model is convolved with rotation
profiles with $v$\,sin$i$=67\,km/s and 68\,km/s, respectively (from Howarth
\etal 1997). It has parameters appropriate for 15\,Mon (see above), being
somewhat cooler than the temperature obtained by Herrero \etal (1992,
\Teff=39\,500\,K). The \iion{S}{IV} emission lines match the observation very
well, both in wavelength and in strength. The supergiant HD\,105056 requires a
much cooler model with much lower gravity (\Teff=31\,600\,K, \logg=3.1,
Pauldrach \etal 1990), but nevertheless we show the spectrum of this star with
the same model in order to emphasize the relative strength of the \iion{S}{IV}
emissions as compared to the main sequence star 15~Mon. The relative strength
of the two emission lines is not matched by the model. This can be caused by
subtle details, like treatment of collisional rates among the levels, and
deserves further investigation.

Our models also include \iion{S}{IV} and  \iion{S}{V} lines in the UV and FUV
spectral regions. They show an \iion{S}{V} absorption line at 1502\AA\ which
supports the identification first suggested by Howarth (1987). In addition, the
models predict an absorption line of \iion{S}{IV} at 1099\AA. It is a multiplet
with four components between 1098.4\AA\ and 1100.05\AA\, arising from
transitions between the lowest $^2$D and $^2$D$^{\rm o}$ terms in the Grotrian
diagram (Fig.\,1). A hitherto unidentified line at this wavelength has been
observed in COPERNICUS spectra of O-stars (Walborn \& Bohlin 1996) and we
suggest the identification with \iion{S}{IV}.

\section{Conclusion}

Finally, sixty years after the first documentation of the 4486,\,4504\AA\
emission lines in O-star spectra, their origin has been revealed; by using
model calculations, we have shown that they are attributable to \iion{S}{IV}.
Since the models are of plane-parallel atmospheres, neither sphericity effects
nor stellar winds need to be invoked for this emission line phenomenon. Future
work may explain the observed strength of these lines as a function of
effective temperature and luminosity class so that, in turn, these selective
emission lines may henceforth be used as a diagnostic tool.

\acknowledgements
We are grateful to Nolan Walborn for useful correspondence and for a preprint of
his paper that initiated this work. We thank Nidia Morrell and Ian Howarth for
generously providing us with their optical data of 15\,Mon and HD\,105056. This
work is supported by DLR (50\,OR\,9705\,5).

\end{document}